\documentstyle[12pt,epsfig]{article}

\begin{document}

\begin{titlepage}
\begin{flushright}
UCLA/98/TEP/1\\
hep-ph/9801337 \\
January, 1998\\
\end{flushright}

\vskip 2.cm

\begin{center}
{\Large\bf What can we learn from $b\rightarrow s \gamma$?}
\footnote{Talk presented at Beauty '97, Fifth International Workshop
on $B$-Physics at Hadron Machines, Los Angeles, CA, October 13-17,
1997.  To be published in Nuclear Instruments and Methods in Physics
Research A.}
\vskip 2.cm

{\large A. K. Grant}

\vskip 0.5cm

{\it  Department of Physics, University of California at Los
Angeles,\\Los Angeles, CA 90095-1547}
\vskip 3cm
\end{center}

\begin{abstract}
We review some recent theoretical progress in the understanding of
weak radiative $B$ meson decay, and discuss the implications of present
and improved measurements of $b\rightarrow s \gamma$ for
supersymmetric models.
\end{abstract}
\vfill
\end{titlepage}

\section{Introduction}

Measurements of flavor changing neutral current processes provide some
of the most stringent experimental tests of the standard model and its
extensions.  In the standard model, the smallness of quantities such
as the $K-\bar{K}$ mixing amplitude is linked to the fact that all such
flavor changing amplitudes vanish at tree level, and can proceed only
at higher orders in the weak interactions.  Because the standard model
predictions are so small, these processes provide an excellent testing
ground for extended models possessing new sources of flavor violation.

Unfortunately, it is not always a simple task to convert the
underlying quark-level weak amplitudes into robust predictions for the
observed hadronic quantities.  Strong interaction effects can
significantly modify the quark model estimates of decay rates.
Luckily, in the case of the $B$ mesons, the large mass $m_b$ of the
$b$ quark makes it possible to systematically compute the effects of
strong interactions in a double expansion in $\alpha_s(m_b)\sim 0.2$
and $\Lambda_{QCD} / m_b$, where $\Lambda_{QCD}\sim 200~{\rm MeV}$ is
an energy scale typical of hadronic quantities such as the hyperfine
splitting in the heavy mesons.  Hence the rare $B$ decays $\bar{B}
\rightarrow X_s \gamma$ and $\bar{B} \rightarrow X_s \ell^+ \ell^-$
with $\ell = e,\mu,\tau$ are especially suitable as tools for testing
the standard model and its extensions.

Over the last year or so, there has been some theoretical progress
pertaining to the inclusive $b \rightarrow s \gamma$ transition.  With
respect to the perturbative calculations, the complete next-to-leading
order (NLO) QCD corrections have been completed, resulting in a
significant reduction of the theoretical error.  In addition, an
interesting new class of non-perturbative corrections to the rate have
been identified.  These effects are of order $\Lambda_{QCD}^2/m_c^2$,
with $m_c$ the mass of the charm quark.  Such effects are potentially
important, of order 10\% or more.  Furthermore, they appear, on the
face of it, to violate the expectations of heavy quark effective
theory, which predicts that all such corrections are suppressed by
inverse powers of the $b$ quark mass.

In the following section, we review the standard model prediction for
the branching ratio $B(\bar{B}\rightarrow X_s \gamma)$.  In Sec.~3, we
discuss the impact of supersymmetry on the rate, and estimate the
range of superpartner masses that might be probed by improved
measurements.  Sec.~4 concludes.

\section{Standard Model Theory}

\begin{figure}
\begin{center}
\epsfig{file=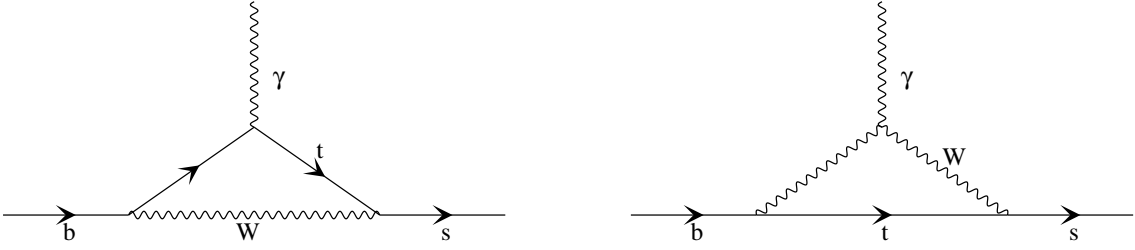,width=6.5in}
\caption{ Feynman diagrams for $b\rightarrow s\gamma$. }
\end{center}
\end{figure}

\subsection{Status of Perturbative QCD calculations}

In the standard model, the $b \rightarrow s \gamma$ decay is mediated
by the $W$ exchange diagrams shown in Fig. 1.  Ignoring QCD
corrections, these diagrams result in an effective Lagrangian of the
form
\begin{equation}
{\cal L}_{b \rightarrow s \gamma} = \frac{G_F e m_b }{4 \sqrt{2}
\pi^2} V_{ts}^{*} V_{tb} C_7 \bar{s}_L \sigma^{\mu\nu} b_R F_{\mu\nu},
\end{equation}
where $C_7$ contains the dependence on the $W$ and top quark masses,
and is given by
\begin{equation}
C_7 = \frac{ 3 x^3 - 2 x^2 }{4(x-1)^4} \ln x - \frac{ 8 x^3 + 5 x^2 -
7 x }{ 24 ( x - 1 )^3 },~~~x = m_t^2/M_W^2.
\end{equation}
Perturbative QCD corrections can be included by use of the
renormalization group equations, which enable us to scale the
coefficient $C_7$ from a high scale $\mu_W\sim M_W$ to a low scale
$\mu_b \sim m_b$.  In this way, the large logarithms of order
$\alpha_s \log(M_W/m_b)$ can be resummed into the coefficient function
$C_7(\mu)$, which now depends on the renormalization scale $\mu$ as
well as the top and $W$ masses.  The leading order QCD corrections
have been known for some time \cite{GSW_M}, and the NLO corrections
have recently been completed \cite{CMM}.  In calculating the branching
ratio, it is convenient to normalize to the semileptonic width.  This
eliminates a large uncertainty arising from the usual factor $m_b^5$
appearing in weak decay rates.  The ratio is given by
\begin{equation}
\label{ratio}
R = \frac{B(\bar{B}\rightarrow X_s \gamma)}{B(\bar{B}\rightarrow X_c e
\bar{\nu}_e)} = \frac{ | V_{ts}^* V_{tb} |^2 }{|V_{cb}|^2} \frac{ 6
\alpha_{QED}}{\pi f(z)} F ( |D|^2 + A )( 1 + \delta_{\rm NP}).
\end{equation}
Here $z=m_{c,{\rm pole}}^2/m_{b,{\rm pole}}^2$, and $f(z)$ is the
phase space factor for the semileptonic width. $F$ includes the QCD
correction to the semileptonic width, while $D$ is the QCD corrected
value of $C_7(\mu)$, and $A$ gives the contribution from certain
infrared logarithms appearing in the $b\rightarrow s\gamma g$ rate.
The full expressions for $f$, $F$, $D$, and $A$ can be found in
Ref.~\cite{CMM}.  Finally, the quantity $\delta_{\rm NP}$ is the the
correction due to non-perturbative binding effects, which we shall
return to below.  The important point here is the dependence on the
renormalization scale $\mu_b$.  Physical quantities should not depend
on $\mu_b$, although at any finite order of perturbation theory, some
residual dependence on $\mu_b$ remains.  Typically $\mu_b$ should be
chosen of order $m_b$, but the variation of the prediction for a range
of $\mu_b\sim m_b/2,2 m_b$ indicates the uncertainty resulting from
higher order corrections.  In leading order, this uncertainty is
roughly 30\%; the inclusion of the NLO corrections reduces this to
about 4\%.

\subsection{Non-Perturbative Corrections}

Apart from the perturbative corrections discussed so far, the rate is
also modified by non-perturbative binding effects.  These effects can
be systematically calculated by means of the heavy quark effective
theory \cite{HQET} together with the operator product expansion (OPE).
Since the energy release in $B$ decays is large compared to the typical
strong interaction scale $\Lambda_{\rm QCD}$, the corrections can be
organized as a series in powers of $\Lambda_{\rm QCD} / m_b$.  The
leading corrections are governed by the quark `pole' mass $m_{b,{\rm
pole}}$, a quantity $\bar{\Lambda}$ that measures the mass of the
light degrees of freedom in the meson, and finally two quantities
$\lambda_{1,2}$ of dimension mass$^2$ that measure the Fermi motion of
the $b$ quark and the hyperfine splitting between the $B$ and $B^*$
mesons, respectively.  $\lambda_1$ and $\lambda_2$ can be expressed in
terms of $B$ meson matrix elements as
\begin{equation}
\lambda_1 = \frac{\langle B | \bar{b} (i D)^2 b | B \rangle}{2
M_B},~~~~ \lambda_2 = -\frac{\langle B | \bar{b} ( g \sigma^{\mu\nu}
G_{\mu\nu} ) b|B\rangle}{12 M_B}.
\end{equation}
The quantities $\lambda_{1,2}\sim \Lambda_{\rm QCD}^2$ and
$\bar{\Lambda}\sim \Lambda_{\rm QCD}$ suffice to compute the
non-perturbative effects through order $1/m_b^2$.  For the
$b\rightarrow s \gamma$ decay, the rate is given by \cite{FLS}
\begin{equation}
\label{HQET_correction}
\Gamma(\bar{B} \rightarrow X_s \gamma) = \Gamma_0 \biggl{(} 1 +
\frac{\lambda_1 - 9 \lambda_2}{2 m_b^2} \biggr{)},
\end{equation}
where $\Gamma_0$ is the quark-level decay rate.  Including the
$1/m_b^2$ contribution to the semileptonic decay rate \cite{MW}, we
find that the ratio $R$ in Eq.~(\ref{ratio}) is enhanced by roughly 1\%
through non-perturbative effects.  Note in particular that when the
width is expressed in terms of the quark pole mass, the leading
corrections are at order $1/m_b^2$ and hence are expected
to be quite small.

\begin{figure}
\begin{center}
\epsfig{file=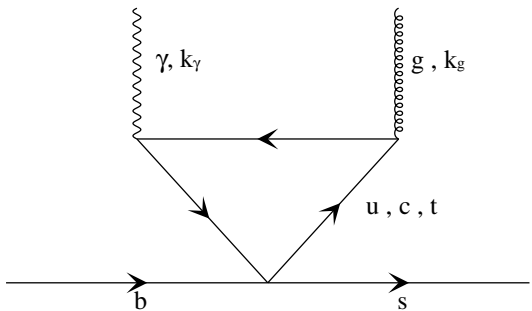,width=3in}
\caption{ Triangle diagram for the $b\rightarrow s \gamma g$ process.
We have omitted a second diagram with the gluon and photon
interchanged. }
\end{center}
\end{figure}

It has recently been shown that the inclusive $\bar{B}\rightarrow X_s
\gamma$ rate is subject to corrections at order $1/m_c^2$
\cite{Voloshin}.  On naive dimensional grounds, these corrections
could be sizable.  Indeed, the scale of these effects is set by
quantities like $\lambda_2/m_c^2\sim 0.08$, so shifts of 10\% or more
are possible.  The ${\cal O}(1/m_c^2)$ contribution to the inclusive
rate comes from an amplitude involving the $b\rightarrow c\bar{c} s$
transition, where the $c\bar{c}$ pair annihilates into the final state
photon after absorbing a soft gluon\footnote{Similar corrections of
this type involving hard gluons have been considered in
Refs.~\cite{KA}}.  The resulting loop diagram is shown in Fig.~2.
Similar loops involving $u$ and $t$ quarks can be safely neglected.
The dependence of this diagram on the charm quark mass can be deduced
using dimensional analysis \cite{Voloshin}.  In terms of the photon
and gluon momenta $k_{\gamma}$ and $k_g$, the diagram depends on the
invariants $k_{\gamma}^2$, $k_g^2$, $k_{\gamma}\cdot k_g$, and $m_c$.
In the limit of an on shell photon and a very soft gluon, all of the
invariants involving four momenta vanish, and we conclude that the
amplitude depends on $m_c$ only.  Explicit evaluation yields a new
contribution to the weak effective Lagrangian which mediates the
$b\rightarrow s \gamma g$ transition:
\begin{equation}
\label{bsgg}
{\cal L}^{b\rightarrow s \gamma g} = -\frac{ e Q_c C_2}{8 \pi^2}
\sqrt{2} G_F V_{cs}^* V_{cb} (\bar{s}_L \gamma^\mu \frac{\lambda^a}{2}
b_L ) \frac{ig}{3 m_c^2} G_a^{\nu \lambda} \partial_\lambda
\tilde{F}_{\mu\nu},
\end{equation}
where $C_2\simeq 1.02$ \cite{CMM} is the coefficient of the
$b\rightarrow c\bar{c}s$ four-Fermi operator.  Including this term in
the OPE for the inclusive $\bar{B}\rightarrow X_s \gamma$ rate, we
find a correction of order $1/m_c^2$ \cite{Voloshin, BIR}:
\begin{equation}
\label{leading}
\frac{\delta \Gamma(\bar{B}\rightarrow
X_s\gamma)}{\Gamma(\bar{B}\rightarrow X_s\gamma)} = -\frac{C_2}{9 C_7}
\frac{\lambda_2}{m_c^2} \simeq +0.025.
\end{equation}
In fact, the approximation leading to Eq.~(\ref{leading}) is not as robust
as one might hope \cite{LRW,GMNP}.  The evaluation of the triangle
diagram involves an integral of the form
\begin{equation}
\label{InvariantFunction}
I(k_\gamma, k_g) = 24 \int_0^1 x\,dx \int_0^{1-x} y\,dy
\frac{1}{\left[ m_c^2 - k_g^2 x(1-x) - k_{\gamma}^2 y ( 1 - y )- 2xy
k_g\cdot k_\gamma \right]},
\end{equation}
which reduces to $1/m_c^2$ in the limit
$k_{\gamma}^2,k_g^2,k_{\gamma}\cdot k_g \ll m_c^2$.  In actuality, we
have $k_{\gamma}^2\equiv 0\ll m_c^2$ and $k_g^2\sim\Lambda_{\rm QCD}^2
\ll m_c^2$; however $k_{\gamma}\cdot k_g \sim m_b\Lambda_{\rm QCD}
\sim m_c^2$.  Hence the result (\ref{leading}), which involves setting
$k_{\gamma}\cdot k_g$ to zero, is potentially subject to large
corrections. These corrections manifest themselves as a series of
higher dimension operators in ${\cal L}^{b\rightarrow s \gamma g}$.
Operators of dimension $(8+n)$ in this series will contribute to the
inclusive rate by an amount proportional to $(m_b \Lambda_{\rm QCD} /
m_c^2)^n$.  Since $(m_b \Lambda_{\rm QCD} / m_c^2) \sim {\cal O}(1)$,
these subleading contributions are not suppressed by any small number.

The first few higher dimension operators in the $b\rightarrow s\gamma
g$ Lagrangian can be found by expanding the charm loop integral in
powers of $k_{\gamma}\cdot k_g/m_c^2$.  The effect of these terms on
inclusive rate can be found as before using the OPE: we have
\cite{GMNP}
\begin{eqnarray}
\frac{\delta \Gamma(\bar{B}\rightarrow
X_s\gamma)}{\Gamma(\bar{B}\rightarrow X_s\gamma)}&=& 
\frac{C_2}{ 9 C_7 } 
\frac{1}{12 M_B} 
\langle B | g \bar{b} \biggl{[}
  \frac{1}{m_c^2} \sigma_{\nu\lambda} G^{\nu\lambda} 
+ \frac{2}{15}\frac{m_b}{m_c^4}
  \sigma_{\nu\alpha} ( i D^{\alpha} )G^{\nu 0} \nonumber\\ &-&
  \frac{3}{700}\frac{m_b^2}{m_c^6}\biggl{(} \sigma_{\nu\lambda} (iD)^2
  G^{\nu\lambda} + \sigma^{\nu\alpha} \{ i D_{\lambda} , i D_{\alpha} \}
  G^{\nu\lambda} \biggr{)} \biggr{]} b | B \rangle.
\end{eqnarray}
Here we have specialized to the rest frame of the $B$ meson, where to
leading order $P_b = ( m_b , {\bf 0})$.  The first term in square
brackets gives the leading contribution of Eq.~(\ref{leading}), while
the subsequent terms give the first two subleading corrections.  In
fact, the first subleading term vanishes at this level of
approximation by virtue of the QCD analogue of the Maxwell equation
${\bf \nabla \times E} = -{\bf \dot{B}}$, together with the heavy
quark equations of motion \cite{GMNP}.  The next subleading term is
certainly negligible, owing to its small $3/700$ coefficient, which
arises from solid angle factors in the photon phase space integration
as well as small coefficients in the Taylor expansion of the loop
integral.  The smallness of these corrections leads us to believe that
the leading $\sim 3\%$ result of Eq.~(\ref{leading}) is in fact
quantitatively correct.

It is interesting to ask whether non-perturbative ${\cal O} ( 1/m_c^2
)$ effects of this type might be important in other $B$ decays
\cite{Rosner_Ciuchini}.  The impact of the charm loop on
$\bar{B}\rightarrow X_s e^+ e^-$ has been estimated \cite{BIR,CRS},
and found to be similar to that in $\bar{B} \rightarrow X_s \gamma$,
of order a few percent.  It is unlikely that corrections of this type
could have a significant impact on the $B$ semileptonic branching
ratio or the ratio of the $B^0$ and $\Lambda_b$ lifetimes.  A
comparison of $\bar{\Lambda}_b \rightarrow X_s \gamma$ with
$\bar{B}\rightarrow X_s \gamma$ is one way to test for spin dependent
effects of this type \cite{Voloshin}.  Since the light degrees of
freedom in the $B$ carry spin 1/2, while those in the $\Lambda_b$
carry spin 0, we expect that the charm loop will have no effect on
$\Lambda_b$ decays.

Finally, we should point out that the $1/m_c^2$ contributions are not
in conflict with heavy quark effective theory.  Indeed, in the HQET limit
$m_b\rightarrow \infty$ with $m_c$ fixed, the corrections are not
proportional to $1/m_c^2$ and have a rather different form
\cite{GMNP}.  The presence of these effects can be traced to a
breakdown of the OPE in processes where the final state photon is
radiated from light quarks, rather than being produced directly in the
$b\rightarrow s$ transition \cite{LRW}.

\subsection{Standard Model Prediction}

Assembling the ingredients of the previous subsections, the standard
model prediction for the branching ratio is \cite{CMM,BKP}
\begin{equation}
\label{prediction}
B(\bar{B}\rightarrow X_s \gamma ) = 3.48 \pm 0.13 \pm 0.28 \times 10^{-4},
\end{equation}
where the first error is due to renormalization scale dependence and
the second is due to errors on the input parameters, which have been
added in quadrature.  A detailed accounting of the errors can be found
in \cite{CMM,BKP}; the dominant errors at present are due to the ratio
of the $b$ and $c$ quark pole masses (5.3\%), the semileptonic
branching ratio (4\%), and the renormalization scale dependence (4\%),
with smaller ( $\sim 3\%$) errors from $\alpha_s(M_Z)$, CKM angles,
and the top quark mass.  One promising way to reduce the theoretical
error is through better determination of the quark pole masses.
Analyses of the lepton energy spectrum or the hadronic invariant mass
spectrum in semileptonic decays have been proposed as means of
extracting the pole masses \cite{FalkB97}.  In addition, measurements
of the photon energy spectrum in $\bar{B}\rightarrow X_s \gamma$ can
be used to provide largely independent constraints on the HQET input
parameters \cite{KL}, including the pole masses.

The prediction (\ref{prediction}) is to be compared with measurements by
CLEO \cite{CLEO} and ALEPH \cite{ALEPH}:
\begin{eqnarray}
B(\bar{B} \rightarrow X_s \gamma) &=& 2.32 \pm 0.57 \pm 0.35 \times 10^{-4}
({\rm CLEO})\\ 
                                  &=& 3.38 \pm 0.74 \pm 0.85 \times 10^{-4}
({\rm ALEPH}).
\end{eqnarray}
For reasons discussed in \cite{FalkB97}, it is important to rely as
little as possible on theoretical models of the photon energy spectrum
in measuring the branching ratio.  Indeed, measurements of the
spectrum can be used to refine our theoretical understanding of the
$B$ system.  As an example, the photon energy spectrum can be related
to the lepton energy spectrum in inclusive semileptonic $b\rightarrow
u$ decays \cite{Neubert}.

\section{$b\rightarrow s \gamma$ and Supersymmetry}

The impact of new physics, and in particular supersymmetry, on flavor
changing neutral currents is the subject of a vast and growing
literature.  The case of $b \rightarrow s \gamma$ has been studied
extensively \cite{SUSY_bsg}.  In this section we review some of the
salient features of supersymmetry as it relates to flavor changing
neutral currents in general and the $b\rightarrow s \gamma$ decay in
particular.

Supersymmetry (SUSY) predicts the existence of a scalar partner for
each standard model fermion, and a fermionic partner for each gauge or
Higgs boson \cite{HK}.  The left- and right-handed fermions have
separate scalar partners, which can have different masses and can mix
with one another.  In addition, softly broken supersymmetry requires
at least one additional Higgs doublet in order to generate masses for
both up- and down-type quarks.  The Higgs spectrum then consists of
two neutral scalars, a neutral pseudoscalar, and a charged Higgs.
When the electroweak symmetry is broken, the charged gauginos and
Higgsinos mix to form two Dirac `charginos,' while the neutral
gauginos and Higgsinos mix to form four Majorana `neutralinos.'

Flavor changing neutral currents are problematic for general
supersymmetric models.  This is a consequence of flavor mixing among
the squarks.  If the squarks are non-degenerate and strongly mixed,
the couplings of the gauginos and Higgsinos will not conserve flavor.
Loops involving (for instance) squarks and gluinos can then give large
contributions to flavor changing processes.  Hence there is a
potential conflict between weak-scale supersymmetry and small flavor
changing neutral currents \cite{sflavor}.

These problems are alleviated if it is assumed that the squarks are
degenerate in mass.  This is the case, for instance, in supergravity
\cite{SUGRA} and gauge-mediated \cite{GM} models of supersymmetry
breaking.  In this case, the gluinos and neutralinos have
flavor-diagonal couplings, while the chargino couplings reduce to the
usual standard model CKM matrix.  Renormalization effects break the
degeneracy of the squarks.  Typically, the squarks will be degenerate
at a high scale ($\sim 10$'s of TeV to $M_{\rm Planck}$, depending on
the model) where supersymmetry breaking is communicated to the
standard model particles.  When the squark masses are renormalized
down to the weak scale, the squarks of the third generation are
lighter than the others.  Even in the presence of this mass splitting,
the gluino, neutralino, and chargino couplings are not strongly
modified in comparison to the case of exact degeneracy.  Tests of
`minimal' models of gauge mediated supersymmetry breaking based on
$b\rightarrow s \gamma$ have been discussed in \cite{Sarid}.

To illustrate the sensitivity of $b \rightarrow s \gamma$ to high mass
superpartners, we will consider a relatively benign scenario with the
following parameters \cite{HW}.  All squarks, except for the lighter
top squark $\tilde{t}_1$, have a common mass of 500~GeV.  The light
stop mass is set to 200~GeV, and is taken to be a maximal mixture of
$\tilde{t}_L$ and $\tilde{t}_R$.  The ratio of the vacuum expectation
values of the two Higgs doublets, $\tan\beta$, is equal to 3, and the
charged Higgs mass is 600~GeV.  Finally, we will assume that the
neutralinos and gluinos have flavor diagonal couplings, while the
chargino couplings are proportional to the standard model CKM matrix.
In this limit, the dominant contributions to $b\rightarrow s\gamma$
arise from loops involving $W$ and the top quark, the charged Higgs
and the top quark, and the charginos and the top squarks.  The charged
Higgs enhances the rate, while the charginos can enhance or suppress
the rate.  Consequently it is not possible to derive model-independent
bounds on superparticle masses, although in non-supersymmetric
two-doublet models, the lower bound on the charged Higgs mass turns
out to be around 340 GeV \cite{Giudice}.  To compute the rate, we must
specify two final parameters: $\mu$, the Higgsino mass parameter, and
$M_2$, the $SU(2)$ gaugino mass.  These parameters determine masses
and couplings of the charginos.  Fig.~3 qualitatively illustrates the
dependence of $B(\bar{B}\rightarrow X_s \gamma)$ on $\mu$ and $M_2$
for this choice of parameters\footnote{We caution the reader that
while Fig.~3 includes all NLO corrections for the standard model case,
the SUSY prediction does not include NLO matching conditions for the
Wilson coefficients.  These have not yet been evaluated.  This
approximation is adequate for our qualitative purposes.}.
\begin{figure}
\begin{center}
\epsfig{file=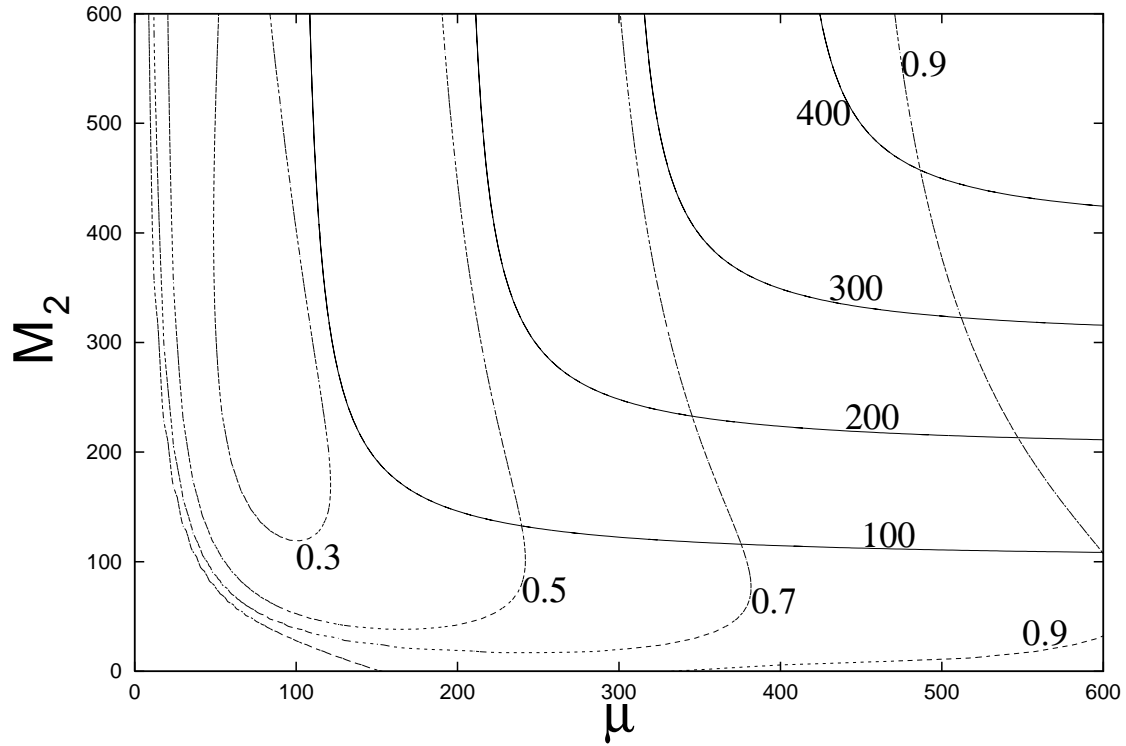,width=4in,angle=-90}
\caption{ Contours of constant chargino mass (solid curves) and
constant $B(\bar{B}\rightarrow X_s\gamma)$ (broken curves).  Both axes
and the chargino masses are in GeV units.  See text for details. }
\end{center}
\end{figure}
For clarity of presentation, we have shown only $\mu>0$; with this
choice of stop mixing, the charginos enhance the rate for $\mu<0$.
The solid curves are contours of constant chargino mass, along which
the lightest chargino has mass 100, 200, 300, and 400~GeV.  The broken
curves are contours along which the ratio $r=B(b \rightarrow s
\gamma)_{\rm SUSY}/B(b \rightarrow s\gamma)_{\rm SM}$ is constant, with
values 0.3, 0.5, 0.7, and 0.9.  We see that deviations from the
standard model at the 10-20\% level are possible for chargino masses
in the 300-400 GeV range.  Consequently, improved measurements will
continue to narrow the allowed region of SUSY parameter space.

\section{Conclusions}

Theoretical understanding of $\bar{B}\rightarrow X_s\gamma$ continues
to improve.  The completion of the NLO QCD corrections has
dramatically reduced the uncertainty on the standard model prediction.
The non-perturbative corrections of order $1/m_b^2$ are well
understood, and the corrections of order $1/m_c^2$, while
unexpected, are both small and calculable.  The sensitivity of
$\bar{B}\rightarrow X_s\gamma$ to new physics is quite remarkable, and
we have reason to hope that improved measurements will continue to add
to our knowledge of short--distance physics.  Improved measurements of
the photon energy spectrum will also be an aid in our understanding of
hadronic physics.

\begin{center}
{\bf Acknowledgments}
\end{center}

It is a pleasure to thank Peter Schlein for organizing an enjoyable
and informative meeting.  I would also like to thank Andrew Morgan,
Shmuel Nussinov and Roberto Peccei for collaboration on the ${\cal
O}(1/m_c^2)$ corrections discussed in Sec.~2.2.  This work was
supported in part by the Department of Energy under Grant No
FG03-91ER40662, Task C.


\begin{thebibliography}{99}

\bibitem{GSW_M} B. Grinstein, R. P. Springer, and M. B. Wise,
Nucl. Phys. B 339 (1989) 269;  M. Misiak, Phys. Lett. B 316 (1991)
127; M. Misiak, Nucl. Phys. B 393 (1993) 23.

\bibitem{CMM} K. Chetyrkin, M. Misiak, and M. M\"unz, Phys. Lett. B
400 (1997) 206, and earlier references therein.

\bibitem{HQET} Heavy Quark Effective theory has been reviewed by in
these proceedings by Mark B. Wise and by Adam F. Falk.

\bibitem{FLS} A. F. Falk, M. E. Luke, and M. J. Savage, Phys. Rev. D 49
(1994) 3367.

\bibitem{MW} J. Chay, H. Georgi, and B. Grinstein, Phys. Lett. B 247
(1990) 399; I. I. Bigi, M. Shifman, N. G. Uraltsev and
A. I. Vainshtein, Phys. Rev. Lett. 71 (1993) 496; A. Manohar and
M. B. Wise, Phys. Rev. D 49 (1994) 4553.

\bibitem{Voloshin} M. Voloshin, Phys. Lett. B 397 (1997) 275.

\bibitem{KA} A. Ali and C. Greub, Phys. Lett. {\bf B361} 46 (1995);
A. Khodjamirian, R. R\"uckl, G. Stoll, and D. Wyler, preprint
WUE-ITP-97-001, MPI-PhT/97-5, Zurich Uni-31/96, Technion-ph-96-24,
hep-ph/9702318.

\bibitem{BIR} G. Buchalla, G. Isidori and S.-J. Rey, preprint
SLAC-PUB-7448 (hep-ph/9705253).

\bibitem{LRW} Z. Ligeti, L. Randall and M. B. Wise, Phys. Lett. B 402
(1997) 178.

\bibitem{GMNP} A. K. Grant, A. G. Morgan, S. Nussinov and
R. D. Peccei, Phys. Rev. D 56 (1997) 3151.

\bibitem{Rosner_Ciuchini} Other effects of charm loop diagrams have
been discussed by Jonathan L. Rosner and by Marco Ciuchini in these
proceedings.

\bibitem{CRS} J.-W. Chen, G. Rupak, and M. J. Savage,
Phys. Lett. B 410 (1997) 285.

\bibitem{BKP} A. J. Buras, A. Kwiatkowski, and N. Pott,
Phys. Lett. B 414 (1997) 157.

\bibitem{FalkB97} These analyses have been discussed by Adam F. Falk
in these proceedings.

\bibitem{KL} A. Kapustin and Z. Ligeti, Phys. Lett. B 355 (1995) 318.

\bibitem{CLEO} CLEO collaboration, M. S. Alam {\it et al.},
Phys. Rev. Lett. 74 (1995) 2885.

\bibitem{ALEPH} ALEPH collaboration, presented at the International
Europhysics Conference on High Energy Physics, Jerusalem, August
19-26, 1997.

\bibitem{Neubert} M. Neubert, Phys. Rev. D 49 (1994) 4623; I.I. Bigi,
M.A. Shifman, N.G. Uraltsev and A.I. Vainshtein, Int. J. Mod. Phys. A 9
(1994) 2467.

\bibitem{SUSY_bsg} See, e.g., S. Bertolini, F. Borzumati, A. Masiero
and G. Ridolfi, Nucl. Phys. B 353 (1991) 591; R. Barbieri and
G.F. Giudice, Phys. Lett. B 309 (1993) 86; V. Barger, M. S. Berger,
P. Ohmann and R. J. N. Phillips, Phys. Rev. D 51 (1995) 2438; A. Ali,
G. F. Giudice and T. Mannel, Z. Phys. C 67 (1995) 417; P. Cho,
M. Misiak and D. Wyler, Phys. Rev. D 54 (1996) 3329; F. Gabbiani,
E. Gabrielli, A. Masiero and L. Silvestrini, Nucl. Phys. B 477 (1996)
321; J. L. Hewett and J. D. Wells, Phys. Rev. D 55 (1997) 5549;
T. Blazek and S. Raby, Indiana University Preprint IUHET-376
(hep-ph/9712257).

\bibitem{HK} For a review, see, e.g., H. E. Haber and G. L. Kane,
Phys. Rept. 117 (1985) 75.

\bibitem{sflavor} S. Dimopoulos and D. Sutter, Nucl. Phys. B 452
(1995) 496.

\bibitem{SUGRA} See, e.g., L. E. Ib\'a\~nez and C. Lop\'ez,
Nucl. Phys. B 256 (1985) 218.

\bibitem{GM} For a reviews see, e.g., A. E. Nelson, University of
Washington preprint UW-PT-97-20, Talk given at 5th International
Conference on Supersymmetries in Physics (SUSY '97), Philadelphia, PA,
27-31 May 1997 (hep-ph/9707442); G. F. Giudice and R. Rattazi,
preprint CERN-TH/97-380 (hep-ph/9801271).

\bibitem{Sarid} E. Gabrielli and U. Sarid, University of Notre Dame
preprint UND-HEP-97-US02 (hep-ph/9707546), and earlier references
therein.

\bibitem{HW} Similar scenarios have been considered in J. L. Hewett and
J. D. Wells, Phys. Rev. D 55 (1997) 5549.

\bibitem{Giudice} M. Ciuchini, G. Degrassi, P. Gambino and
G. F. Giudice, preprint CERN-TH/97-279 (hep-ph/9710335).


\end{thebibliography}
\end{document}